\documentclass[preprint,12pt]{aastex}
\def\lsim{\lower.5ex\hbox{$\; \buildrel < \over \sim \;$}}
\def\gsim{\lower.5ex\hbox{$\; \buildrel > \over \sim \;$}}
\def\abeq{\lower.7ex\hbox{$\; \buildrel \sim \over - \;$}}

\def\t{\ifmmode {\tau} \else $\tau$ \fi}

\def\ref{\noindent \hangafter=1 \hangindent=0.7 truecm}

\def\cm{\ifmmode {\rm cm}^{-1} \else cm$^{-1}$ \fi}
\def\s{\ifmmode {\rm s}^{-1} \else s$^{-1}$ \fi}
\def\cc{\ifmmode {\rm cm}^{-3} \else cm$^{-3}$ \fi}
\def\cs{\ifmmode {\rm cm}^{-2} \else cm$^{-2}$ \fi}
\def\g{\ifmmode \gamma \else $\gamma$\fi}
\def\G{\ifmmode \Gamma \else $\Gamma$\fi}

\def\kms{\ifmmode {\rm km\ s}^{-1} \else km s$^{-1}$\fi}

\usepackage{graphicx}

\begin{document}

\title{Optical Polarimetry of the Jets of Nearby Radio Galaxies:  I. The Data}

\author{Eric S. Perlman\altaffilmark{1,2}, C. A. Padgett\altaffilmark{1},
Markos Georganopoulos\altaffilmark{1,3}, William B. Sparks\altaffilmark{4}, 
John A. Biretta\altaffilmark{4}, Christopher P. O'Dea\altaffilmark{4,5},
Stefi A. Baum\altaffilmark{4,6}, Mark Birkinshaw\altaffilmark{7},
D. M. Worrall\altaffilmark{7}, Fred Dulwich\altaffilmark{7},
Sebastian Jester\altaffilmark{8,9}, Andr\'e Martel\altaffilmark{2},
Alessandro Capetti\altaffilmark{10}, J. Patrick Leahy\altaffilmark{11}}

\altaffiltext{1}{Department of Physics, Joint Center for Astrophysics,
University of Maryland-Baltimore County, 1000 Hilltop Circle, Baltimore, MD
21250, USA.  E-mail:  (perlman, apadgett, markos)@jca.umbc.edu}

\altaffiltext{2}{Department of Physics and Astronomy, Johns Hopkins University,
3400 North Charles Street, Baltimore, MD 21218, USA}

\altaffiltext{3}{Laboratory for High-Energy Astrophysics, NASA's Goddard
Space Flight Center, Code 660, Greenbelt, MD  20771}

\altaffiltext{4}{Space Telescope Science Institute, 3700 San Martin Drive,
Baltimore, MD 21218, USA.  E-mail: (sparks, biretta)@jca.umbc.edu}

\altaffiltext{5}{Current Address: Department of Physics, Rochester Institute of
Technology, 84 Lomb Memorial Drive, Rochester, NY  14623-5603 USA.  E-mail:
odea@cis.rit.edu}

\altaffiltext{6}{Current Address:  Center for Imaging Science, Rochester
Institute of Technology, 54 Lomb Memorial Drive, Rochester, NY 14623-5604
USA. E-mail: baum@cis.rit.edu}

\altaffiltext{7}{H. H. Wills Physics Laboratory, University of Bristol, Tyndall Avenue,
Bristol  BS8 1TL, UK.  E-mail: (Mark.Birkinshaw, d.worrall,
Fred.Dulwich)@bristol.ac.uk}

\altaffiltext{8}{Fermilab, MS 127, PO Box 500, Batavia, IL 60510, USA}

\altaffiltext{9}{Current Address:  School of Physics and
Astronomy, University of Southampton, Highfield, Southampton SO17 1BJ, UK. 
E-mail:  jester@phys.soton.ac.uk}

\altaffiltext{10}{ INAF-Osservatorio Astronomico di Torino, I-10025 Pino
Torinese, Italy.  E-mail:  capetti@to.astro.it}

\altaffiltext{11}{Jodrell Bank Observatory, School of Physics \& Astronomy, The
University of Manchester, Macclesfield, Cheshire SK11 9DL, UK.  E-mail: 
jpl@jb.man.ac.uk}

\begin{abstract}

In this paper, the first in a series, we present an overview of new {\it Hubble
Space Telescope} ({\it HST}) imaging polarimetry of six nearby radio galaxies
(3C 15, 3C 66B, 3C 78, 3C 264, 3C 346, and 3C 371) with optical jets. These
observations triple  the number of  extragalactic jets with
subarcsecond-resolution optical polarimetry. We discuss the polarization  
characteristics of each jet and,  as our Stokes {\it I} images also represent
by far the deepest optical images yet obtained of each of these jets,  we also
discuss the morphology in total flux of each jet in detail. We find evidence of
high optical polarization, averaging $20\%$, but reaching upwards of $\sim
50\%$ in some objects, confirming that the optical emission is synchrotron, and
that the components of the magnetic fields perpendicular to the line of sight
are well ordered.  We find a wide range of polarization morphologies, with 
each jet having a somewhat different relationship between total intensity and
polarized flux and the polarization position angle.  We find two trends in all
of these jets.  First,   jet ``edges'' are very often associated with  high
fractional optical polarizations, as also found in earlier radio observations
of these and other radio jets.    In these regions, the magnetic field vectors
appear to  track the jet direction, even at bends, where we see particularly
high  fractional polarizations.  This indicates a strong link between the local
magnetic field and jet dynamics.  Second, optical flux maximum regions are
usually well separated from maxima in fractional polarization and often are
associated with polarization minima. This trend is not found in radio data and
was  found in our optical polarimetry of M87 with HST.   However, unlike in
M87, we do not find a general trend for near-90$^\circ$ rotations in the
optical polarization vectors near flux maxima. We discuss possibilities for
interpreting these trends, as well as implications for jet dynamics, magnetic
field structure and  particle acceleration. 

\end{abstract}

\maketitle

\section{Introduction}

Relativistic jets have been known for many years as a hallmark of powerful
extragalactic radio sources. The compilation of Liu \& Zhang (2002) listed over
660 sources with known, resolved radio jets, as of December 2000. The number of
optical and X-ray jets is far smaller: about three dozen are known in each band
(see the WWW pages maintained by Harris (XJET; Harris 2003) and Jester (Jester
et al. 2004) respectively for lists of confirmed X-ray and optical jets). This
does not signify a paucity of optical and X-ray emission, but instead, reflects
the fact that few comprehensive surveys for jets have been done in those
bands.  Indeed, until the advent of the {\it Hubble} Space Telescope ({\it
HST}) and {\it Chandra} X-ray Observatory, no survey had been done which  was
reasonably optimized for the detection of jets.  Even now, there have been few
{\it HST} surveys for jets: the largest, the 3CR snapshot surveys (optical: de
Koff et al. 1996, Martel et al. 1999; UV: Allen et al.  2002; near-IR: Madrid
et al. 2006), includes more than 100 objects (not all of which have radio
jets), but is complete only to $\sim 21$ mag/ arcsec$^2$ due to the snapshot 
nature of the survey (typical integration time was $\sim 300$ seconds with
WFPC2). This is far less deep than required to see, for example, the knots in
the jet of PKS 0637-752 (Chartas et al. 2000, Schwartz et al. 2000).
Nevertheless, the WFPC2 survey achieved an overall success rate of 13\%  for
jet detection (Martel et al. 1999), and additional objects were discovered with
NICMOS (e.g., Chiaberge et al. 2005, Floyd et al. 2006). More comprehensive and
specifically targeted surveys have been  done in the X-rays, by Sambruna et al.
(2002, 2004) and Marshall et al. (2005), with the former of these also having a
deeper (full orbits with ACS) {\it HST} component.  These surveys have  60-70\%
success rates of finding X-ray and optical jet emission from quasars with known
radio jets.  Thus it is likely that a large proportion (perhaps even a
majority) of jets generate significant levels of both optical and X-ray
emission.

This explosion in the number of known optical and X-ray jets has been
accompanied by a large increase in our knowledge about these objects. And yet,
despite this, we are sorely lacking in information regarding the structures and
composition of jets, as well as how these elements interact to produce the
energetic emissions we see.  Part of the reason we lack such basic information
is that it is difficult to obtain: since AGN jets are ionized flows, we lack
many of the diagnostics that can be used to decipher physical conditions in
other object classes. For sources that emit synchrotron radiation, polarimetry
is a powerful tool that can yield direct information on the interactions
between fields and particles.  The polarization vectors of synchrotron
radiation indicate the magnetic field direction in the emission region, while
the fractional polarization reflects the relative ordering of the magnetic
field.  Since the electrons that emit synchrotron radiation in the optical and
radio wavebands differ in Lorentz factor (and hence energy and cooling lengths)
by factors of more than 100, comparing polarimetry in the radio and optical
bands gives information on the magnetic field in regions occupied by particles
in very different energy regimes.  Contrary to what one would expect in a
homogeneous jet, polarimetry of the M87 jet shows different characteristics of
polarization in the radio and optical bands (Perlman et al. 1999): in the
optical, there is a strong anti-correlation between flux and polarization, and
at the upstream ends of knots the optical magnetic field vectors become
perpendicular to the jet. Much smaller changes are seen in the radio.  Thus
the optical and radio emission in the M87 jet must
occur in somewhat different regions.  In Perlman et al. (1999) we interpreted
this as evidence that higher energy particles are concentrated closer to the
center of the jet flow, with shocks that accelerate electrons and compress
magnetic field lines located in the jet interior.

While {\it HST} imaging observations now exist for virtually all known optical
jets, it has been much more difficult to obtain {\it HST} polarimetry for these
objects.  Previous to this work, {\it HST} polarimetry existed for only three
optical jets:  M87 (Pre-COSTAR FOC: Boksenberg et al. 1992, Thomson et al.
1995, Capetti et al. 1997, WFPC2:  Perlman et al. 1999), 3C 273 (Pre-COSTAR
FOC: Thomson et al. 1993) and 3C 293 (NICMOS:  Floyd et al. 2005).  Of these
three, only in M87 were the observations of sufficient S/N ($\gsim 30$) to
track accurately both the polarization fraction and orientation throughout
virtually the entire jet.  The other two objects had much lower S/N (typically
$\sim 10-15$ for 3C 293 and $\lsim 4-20$ for 3C 273), making conclusions based
on these quantities much less certain.   As a result very little is known about
the energetic structures of most optical jets, including the extent to which 
the radio and optical emitting electron populations are co-spatial and
encounter a common magnetic environment, as well as the role played by the
magnetic field structure in the acceleration of high-energy particles.

We are engaged in a project to obtain {\it HST} polarimetry of the nearest and
brightest optical jets, in  order to remedy this situation.  In this paper, we
present polarimetric observations of six nearby jets. These objects, namely 3C
15, 3C 66B, 3C 78, 3C 264, 3C 346 and 3C 371, constituted the six brightest
optical jets as of 1999, when the project was first proposed for HST
observations.  This paper presents the {\it HST} observations in an atlas form,
as well as a qualitative discussion of the polarization properties of
individual jets and sample properties.  Upcoming papers will concentrate on
individual jets and physical interpretation  (for example, 3C 15:  Dulwich et
al. 2006; 3C 66B:  Perlman et al. in preparation; 3C 264:  Padgett et al., in
preparation), including multiwavelength data as well as the relationship of
polarized emissions to the spatially resolved optical and broadband structure
and spectrum of each jet (e.g., for M87, Perlman et al. 1999, 2001, Perlman \&
Wilson 2005).  

The paper is laid out as follows. In \S 2 we discuss the observations and data
reduction procedures. In \S 3, we detail the results on each object.  In \S 4,
we compare them to the few objects for which previous {\it HST} polarimetry has
been done, and present the first discussion of the sample properties of jets in
polarized optical light.  We close our discussion in \S 5 with a summary.
Throughout the paper we assume a cosmology with $H_0=70 {\rm ~km ~s^{-1}
~Mpc^{-1}}$ (Riess et al. 2005 and references therein), $\Omega_{m,0}=0.3$ and
$\Omega_{\Lambda, 0}=0.7$ (Riess et al. 2004; Bennett et al. 2003; Perlmutter et al.
1999). 

\section{Observations and Data Reduction}

Polarimetric observations were obtained with {\it HST}. In Cycle 10 (program
GO-9142), we used WFPC2 to observe 3C 78 and 3C 264, while in Cycle 12 (program
GO-9847), ACS was used to observe 3C 15, 3C 66B, 3C 346 and 3C 371. Details are
given in Table 1. Below, we discuss the observing setups and data reduction
particulars.

\begin{deluxetable}{cccccc}
\tablecolumns{5}
\tablewidth{0pt}
\tablecaption{Observations}

\tablehead{
\colhead{Object} &
\colhead{Instrument} &
\colhead{Filters} &
\colhead{Date} &
\colhead{Integration Time(s)}}

\startdata

3C 15 & ACS (WFC)  & F606W+POL0V  & 9 Dec 2003  & 2872 \\
3C 15 & ACS (WFC)  & F606W+POL60V & 9 Dec 2003  & 2872 \\
3C 15 & ACS (WFC)  & F606W+POL120V& 9 Dec 2003  & 2872 \\
\\
3C 66B& ACS (WFC)  & F606W+POL0V  & 6 Aug 2004  & 2968 \\
3C 66B& ACS (WFC)  & F606W+POL60V & 6 Aug 2004  & 2968 \\
3C 66B& ACS (WFC)  & F606W+POL120V& 6 Aug 2004  & 2968 \\
\\
3C 78 & WFPC2 (PC) & F555W+POLQ$^a$   & 27 Aug 2001 & 13500 \\ 
3C 78 & WFPC2 (PC) & F555W+POLQ$^a$   & 7 Oct 2001  & 13500 \\
3C 78 & WFPC2 (PC) & F555W+POLQ$^a$   & 7 Nov 2001  & 13500 \\
\\
3C 264& WFPC2 (PC) & F555W+POLQ$^a$   & 6 Mar 2003  & 6500 \\
3C 264& WFPC2 (PC) & F555W+POLQ$^a$   & 25 Apr 2003 & 6500 \\
3C 264& WFPC2 (PC) & F555W+POLQ$^a$   & 1 Jun 2003  & 6500 \\
\\
3C 346& ACS (WFC)  & F606W+POL0V  & 19 Aug 2003 & 502 \\
3C 346& ACS (WFC)  & F606W+POL60V & 19 Aug 2003 & 502 \\
3C 346& ACS (WFC)  & F606W+POL120V& 19 Aug 2003 & 502 \\
\\
3C 371& ACS (WFC)  & F606W+POL0V  & 20 Sep 2003 & 2828 \\
3C 371& ACS (WFC)  & F606W+POL60V & 20 Sep 2003 & 2828 \\
3C 371& ACS (WFC)  & F606W+POL120V& 20 Sep 2003 & 2828 \\
3C 371& ACS (WFC)  & F606W+CLEAR$^b$& 20 Sep 2003 & 428 \\
\enddata

\tablenotetext{a}{As the polarizer does not rotate, to obtain observations at
three different position angles,  it was necessary to obtain observations on
three different days (see \S 2.1).}

\tablenotetext{b}{Observations obtained to ensure polarization calibration
as well as to measure accurately the PSF.  See text for details.}

\end{deluxetable}

\subsection{Observing Programs}

Because of the different characteristics of each instrument, it
was necessary to adopt different observing strategies for the two programs. 
With WFPC2, we used the combination of the POLQ and F555W filters.  This 
combination was chosen for maximum throughput of parallel-polarized radiation
and minimal transmission of cross-polarized light.   Given the small size of
the two jets observed with WFPC2, as well as concerns about physical
resolution, we chose to use the PC chip, rather than the WF chips as had been
used for M87 by Perlman et al. (1999).  This necessitated long integrations and
reduced the unvignetted field of view to $\sim 16'' \times 16''$.    The POLQ
filter can rotate through only $\sim 50^\circ$, and all
of the rotated positions leave most or all of the PC filter either vignetted or
not covered by the same polarizer.  For that reason, it was necessary to fix
the rotation of the POLQ filter at $0^\circ$.  Obtaining images on the PC at
three position angles (necessary to reconstruct the Stokes Parameters), then
necessitated observations at three different epochs during which ORIENTs of
$\sim \theta_0 + 0^\circ, 60^\circ$ and $120^\circ$ (where $\theta_0$ is the
ORIENT in epoch 1,  and windows of $\pm 15^\circ$ were allowed for ease of
scheduling) respectively were used.  ORIENTs were also chosen carefully to
avoid the possibility of the jet lying along a diffraction spike (in some cases
this meant deviating slightly  from the ideal set of orientations).  In order
to minimize the effects of variability, we further required that observing
epochs be less than 2 months from one another.  Sub-pixel dithering was used,
enabling us to eliminate hot pixels and maximize resolution.

The particulars of the ACS are much more favorable to polarimetry
observations.  Not only is its sensitivity much higher, but also it has
polarization filters optimized for both the UV and optical bands, and each
polarizer set has filters at $0^\circ, 60^\circ$ and $120^\circ$ that cover the
same field of view.  This latter fact eliminated the need for multiple
observing epochs, while the higher quantum efficiency allowed much shorter
integration times for objects that were considerably fainter (by 1-2.5
mag/arcsec$^2$) than those observed with WFPC2 -- indeed fewer orbits were
necessary for the ACS program (15 as opposed to 18) despite including twice the
number of objects.  With ACS, we chose the WFC array to   minimize observing
time.  Since approximately 3/4 of the WFC field is vignetted with the
polarizers, we used the WFC1-1K sub-array, resulting in a $35''$ field of
view.  Despite the reduction in resolution as compared to the HRC chip, this
was necessary given the faintness of the jets in our Cycle 12 program (between
19-20 mag/arcsec$^2$ compared to 18 mag/arcsec$^2$ for our Cycle 10 targets). 
As with the WFPC2 program, ORIENTs were chosen carefully to avoid the
possibility of the jet lying along a diffraction spike.  This was critical for
our observations of 3C 371, a very bright BL Lac object where the central PSF
was heavily saturated.   As with the WFPC2 observations, sub-pixel dithering
was used, allowing us to eliminate hot pixels and maximize resolution.  For one
object (3C 371) observations without the polarizers were also obtained (on the
advice of ACS instrument scientists) to check the calibration.

\subsection{Data Reduction} 

The {\it HST} data were recalibrated using the best available flat fields and
dark count images within STSDAS using standard techniques (e.g., Holtzmann et
al. 1995).  Each ACS image set was combined using multidrizzle (Koekemoer et
al. 2004), incorporating the best available corrections for the chip geometries
(Anderson \& King 2004).  The WFPC2 image sets were combined using interactive
methods in STSDAS, also incorporating the best available distortion corrections
for WFPC2 (Anderson \& King 2003). To register and rotate to the north=up,
east=left orientation, we used cross-correlation techniques for all images. For
the WFPC2 datasets, this was considerably more difficult because of the
undersampling of the PSF and its significant changes over the PC, combined with
the geometric distortions in the PC and the fact that it was necessary to allow
the HST to rotate under the source to obtain all three polarizer observations.
In practical terms, this meant that in the WFPC2 datasets we had to mask out
the  jet and nucleus in cross-correlation (thus relying on off-center globular
clusters for the signal), while in the ACS datasets we were able to use the
entire image.  We tested the correctness of the registration by shifting
slightly the position of the individual images along the x and y axes; for the
WFPC2 data this procedure yielded errors of about $\pm 0.15$ pixels (i.e., $\pm
0''.068$), but for the ACS data the errors were much smaller ($\pm 0.05$
pixels, i.e., $\pm 0''.025$). Once the images were registered, galaxy
subtraction was performed on each polarizer image (see below). 

To increase signal to noise, the ACS image sets were also convolved with a
$\sigma = 1$ pixel circular Gaussian, then binned to $ 0''.1/$pixel (i.e.,
twice the native pixel scale), whereas the WFPC2 image sets were left at
$0''.0455$/pixel, but convolved with the same Gaussian.  The WFPC2 datasets
were then combined using formulae from the WFPC2 Polarization Calibration Tool
(Biretta \& McMaster 1997), which yields Stokes  $U$, $Q$ and $I$ that are
combined in a standard way to produce emission weighted
fractional polarization (defined as $P=(Q^2+U^2)^{1/2}/I$) and apparent
magnetic field position angle (defined as MFPA = $1/2 \times \tan^{-1}(U/Q)+
90^\circ$) images.   The ACS data were combined using the prescription given
in the ACS Data Handbook (Pavlovsky et al, 2005) to yield Stokes parameter,
fractional polarization and MFPA images.

The well-known Rician bias in $P$ (Serkowski 1962) was accounted for in both
the ACS and WFPC2 data, using a Python code adapted from the STECF IRAF package
(Hook et al. 2000).  This code debiases the $P$ image following Wardle \&
Kronberg (1974), and calculates the error in polarization PA, accounting for
the non-Gaussian nature of its distribution (see Naghizadeh-Khouei \& Clarke
1993). In performing this calculation, pixels with signal to noise $(S/N)< 0.5$
were excluded outright, and since the debiasing is done with a ``most-probable
value'' estimator, pixels where the most-probable value of $P$ was negative, or
above the Stokes $I$ value ({\it i.e.} $P > 100\%$) were blanked. Rician bias
was not accounted for in earlier {\it HST} polarimetry reduction 
procedures.  The results of Perlman et al. (1999) on M87 should not be
significantly affected by Rician bias, thanks to their high S/N;
however, it is likely that earlier observations of M87 and/or 3C273 were
significantly affected due to their much lower S/N. 

As a test of our polarimetric calibration, we used aperture photometry to
measure the fractional polarization of both the galaxy light and globular
clusters.  All the results were consistent (within the uncertainties) with
unpolarized emission, as one would expect, with a typical standard deviation of
3\%.  We also compared the flux of individual jet components derived from the
Stokes I image of 3C 371 with those derived from the observations done without
the polarization filters.  Those values matched to within the uncertainties as
well.

Galaxy subtraction was done using the IRAF tasks ELLIPSE, BMODEL and IMCALC in
STSDAS. To successfully model each galaxy, it was necessary to mask out the
jet, stars and globular clusters in each image set. In addition to this, a
single $3\sigma$ rejection was done for each isophote to account for noise.
This method is of necessity an iterative process, as fainter features do not
become apparent until the initial iteration of galaxy subtraction is complete.
Failure to mask out a globular cluster would produce a circular ``ringing''
centered at the distance of the cluster. 

Specific image features required us to take extra care with most of these
objects. In 3C 66B, 3C 78 and 3C 264, the presence of dust rings with nuclear
isophote anomalies (Sparks et al.  2001, Baum et al. 1997) required masking
and/or interpolation over these features, and in the case of 3C 264 it was also
necessary to interpolate to properly fit out the ``ring''. These rings tend to
produce deep negative contours in our resulting frames, which represent an
over-subtraction. The error in the isophotal fits in these regions reflect the
fact that the galaxy model is incorrect, and this error has been propagated
through the reduction (see below). In the case of 3C 346, the presence of a
near neighbor galaxy (within $5''$) required us to mask the entire
south-east portion of the image to model the host galaxy, followed by masking
out the jet and unsubtracted regions to model the companion.  Finally, in the
case of 3C 371, the only object where saturation was a significant issue, the
most successful model was produced by median combining images rotated by
$0^\circ$, $90^\circ$, $180^\circ$, and $270^\circ$ about the centroid of the
central point source. Due to the slightly non-circular nature of the source,
this produced slightly over-subtracted regions in the north-west, and
south-east quadrants, and slightly under-subtracted regions in the other two.
Tiny-Tim PSF subtraction, followed by isophotal galaxy subtraction, suffered
from severe residuals due to saturation and the imprecisions associated with
the Tiny-Tim PSF.

\subsection{Uncertainties}

Gaussian error propagation was used to compute errors in the Stokes $I$, $Q$,
and $U$ images. The propagated errors include Poisson errors, an additive noise
term (the rms background calculated post galaxy subtraction) and rms errors in
the isophotal fits. The resulting uncertainties in the Stokes $Q$ and $U$
images are approximately Gaussian in nature, with their values being
approximately equal to the sum in quadrature of the individual polarizer image
errors. Hence, Gaussian error propagation for $P$ is appropriate for our
purposes. The $1 \sigma$ confidence limits for MFPA were calculated as
mentioned in \S 2.2. 

The two objects observed with the WFPC2/PC (3C 78 and 3C 264) are subject to
somewhat greater uncertainties than those observed with ACS, due to two
factors. The first is the lower sensitivity of the WFPC2, combined with its 
lower gain (7$e-$/ADU versus 1 $e-$/ADU for the ACS),
which are, however, ameliorated by the much longer observing times.  The second,
more significant issue is that, as noted by Perlman et al. (1999), we must 
account for the uncertainty in registering the images. This effect is
negligible in the ACS data sets because the ACS exhibits smaller PSF variations
than does the PC;  moreover, the roll-angle differences were insignificant,  so
the registration was not in question. For the PC image sets, a $\pm 0.15$ pixel
de-registration along the direction of the jet will have an effect of $\delta P
\lsim 0.05$ and $\delta PA \lsim 5^\circ$ in most jet regions.  This effect is
somewhat less critical for these sources than it was for M87 due to our use of
the PC, which better samples large gradients in surface brightness, as well as
their particular morphologies, which have rather smooth jets (with few sudden,
bright features in the jet or galaxy) and in addition are also rather short
(less than $3''$), the latter ensuring that for most jet regions, any effect
due to registration is not dominant although its contribution is not
negligible. 

The polarimetric calibration of the ACS polarizers for the WFC is not yet
complete.  The main problem that remains is to account for the ``ripples'' in
the polaroid film used in the ACS polarizers.  These ripples add to the 
geometric distortion of the images, and cause the wings of the PSF to vary
across the chip and between polarizers, leading to apparently high
polarizations at the edges of unpolarized point sources due to the  high
internal polarization of the ACS.   Kozhurina-Platais \& Biretta (2004) have
characterized this distortion and its effect on polarimetry for HRC
observations; however, a similar work has yet to be completed for the WFC,  for
which the instrumental polarization has not yet been fully constrained. 
Following the ACS Data Handbook (Pavlovski et al. 2005), we include a 10\%
error in the fractional polarization (i.e., this would introduce an extra 2\%
error term in a source that was 20\% polarized) and a $3^{\circ}$ error in the
MFPA of our ACS data, to account for both of these calibration problems.  This
is in addition to the propagated errors discussed above.


\section{Results}

Our results are shown in Figures 1-7.  Each of these shows one or more views of
the Stokes I image; where necessary, we include views with both high and low
contrast as well as multiple fields of view.  We give two views of the
polarimetry of each jet, both in fractional polarization overlaid with flux
contours, and as a polarization vector image with flux contours.  Each image is
in a north-up, east-left configuration, and all images have been galaxy
subtracted.  We have clipped the polarization maps at the 3 $\sigma$ level.  As
can be seen, these jets display a broad variety of morphologies, both in total
intensity as well as polarization. Three jets (3C 15 [Figure 1], 3C 346 [Figure
6] and 3C 371 [Figure 7]) are rather knotty, while three others (3C 66B
[Figures 2, 3], 3C 78 [Figure 4], and 3C 264 [Figure 5]) are much smoother.  

 We give background information on these sources, as well as all other jets
imaged polarimetrically with {\it HST}, in Table 2.  The columns in Table 2
are, in order:  (1) The source name; (2) its redshift; (3) the total power at
178 MHz (a measure of the power put out by the entire source; at this low
frequency it is usually dominated by the lobes); (4) The FR type (Fanaroff \&
Riley 1974); (5) and (6) The length of the jet as seen in the radio and optical
(in kpc); (7) The mean surface brightness (in R band); (8) References and
notes; (9) Average polarization in the jet.

\begin{deluxetable}{l c c c c c c c c}
\tablecolumns{9}
\tablecaption{Jets with HST Polarimetry} 
\tablewidth{0pt}

\tablehead{
\colhead{Source} & \colhead{$z$} &  \colhead{$\log P_{178}$} & \colhead{FR} & 
\colhead{$L$(rad)} & \colhead{$L$(opt)} & \colhead{$\mu_{R,ave}$} &
\colhead{Ref./} & \colhead{Avg.} \\

&  & \colhead{W/Hz} & \colhead{Type}  & \colhead{kpc}& \colhead{kpc}&
\colhead{mag/arcsec$^2$}&\colhead{Notes}&\colhead{\%P}}

\startdata
M87    & 0.0044  & 25.6 & I    & 89  & 2.0 & 16.3 & 1,2,3 & 35 \\
3C 15  & 0.0730  & 26.2 & I/II & 66  & 8.4 & 20.4 & 4     & 20 \\
3C 66B & 0.0213  & 25.4 & I    & 153 & 2.7 & 20.3 & 4     & 20 \\
3C 78  & 0.0287  & 25.5 & I    & 129 & 1.0 & 18.9 & 4     & 12 \\
3C 264 & 0.0217  & 25.4 & I    & 41  & 1.1 & 17.9 & 4     & 25 \\
3C 273 & 0.1583  & 27.6 & II   & 62  &  62 & 20.8 & 5     & $\sim$10 \\
3C 293 & 0.0450  & 25.0 & I/II & 210 & 1.5 & 24.1 & 6,7   & $\sim$6 \\
3C 346 & 0.1620  & 26.8 & II   & 37  & 7.4 & 18.3 & 4     & 23 \\
3C 371 & 0.0510  & 25.3 & I    & 284 & 4.0 & 20.4 & 4     & 21 \\
\enddata
\tablenotetext{1}{Boksenberg et al. (1992); pre-COSTAR FOC, low S/N and incorrect registration.}
\tablenotetext{2}{Capetti et al. (1997); pre-COSTAR FOC.}
\tablenotetext{3}{Perlman et al. (1999); used WFPC2/WF so not full resolution.}
\tablenotetext{4}{This paper.}
\tablenotetext{5}{Thomson et al. (1993); pre-COSTAR FOC and low S/N in most regions.}
\tablenotetext{6}{Floyd et al. (2005); NICMOS; radio jet is two-sided but only E jet is seen in optical.}
\tablenotetext{7}{Polarimetry was done in K-band because Jet and nucleus are highly extincted.}

\end{deluxetable}


{\it 3C 15.}  3C 15 (Figure 1) is a radio galaxy at a redshift $z=0.073$.   At
radio frequencies, it shows diffuse emission from extended radio lobes, a
bright, one-sided jet as well as a weak counter-jet (Leahy et al. 1997).  Its
overall radio power places it at the low end of the FR II regime (i.e., $L_{178
{\rm MHz}} \gsim 10^{25} {\rm ~W ~Hz^{-1}}$, Fanaroff \& Riley 1974), while its
edge-dimmed, diffuse morphology is more characteristic of FR I galaxies (as
opposed to FR IIs where prominent hotspots are more commonly seen).  Its
optical jet was first discovered with {\it HST} observations (Martel et al.
1998), and X-ray observations with {\it Chandra} were discussed by Kataoka et
al. (2003).  At 8.4 kpc long, its optical jet is over four times longer in 
projected size than that of M87 (Sparks, Biretta \& Macchetto 1996).

The jet emission is resolved into four knot regions on radio maps (Leahy et al.
1997), three of which can be seen in the optical (Martel et al. 1998).  The
{\it HST} also resolves the innermost of the radio emission regions into two
fairly distinct knots.  Here we follow the scheme of Martel et al. (1998) in
referring to the emission regions in the 3C 15 jet, and label them  A through
D, respectively, outwards from the nucleus. X-ray emission is seen only from
two knots (A and C; Kataoka et al. 2003, Dulwich et al. 2006). The jet's
optical polarization averages about 20\%.  The highest optical polarizations
are seen along the outer edges (specifically away from knot maxima) of the knot
A-B complex ($\approx 30\%$), as well as at the ends of knot C (35-40\%). 
Reduced polarizations are seen in a curved region that extends throughout the
A-B complex, with minima near the flux maxima of knots A and B ($\approx 0$ and
$5\%$, respectively).   The predominant MFPA seen in the optical in the A-B
complex is parallel to the jet, but significant rotations are seen near the
flux maximum of A as well as in a region inclined to the jet direction through
B and at the downstream end of B.  Further out, in knots C and D, the
predominant MFPA we see is perpendicular to the jet.  In Dulwich et al. (2006),
we interpret these changes as evidence of a helical twist in the magnetic
field  in knots A and B, followed by a strong shock at knot C.  

{\it 3C 66B.} 3C 66B (Figures 2 and 3) is the nearest of our jet sources, at a
redshift of $z=0.0213$ (Stull et al. 1975).  Its optical counterpart is an
elliptical galaxy associated with a small group in the vicinity of the cluster
Abell 347.  The optical galaxy features a dust disk, at a radius of $0''.5$, as
was noted by Sparks et al. (2000); evidence of this feature can also be seen on
our images (Figure 2).  Its radio structure is a hybrid between an
edge-darkened double and a head-tail morphology (Leahy, J\"agers \& Pooley
1986; Hardcastle et al. 1996).  A two-sided inner radio jet extending over
$25''$ is seen, with both jets curving towards the East at distances $>20-30''$
from the core.  The optical jet emission was first detected by Butcher, van
Breugel \& Miley (1980), and further ground-based work (Fraix-Burnet, Nieto \&
Poulain 1989) revealed the presence of five distinct knots (for which we use
the terminology of Hardcastle et al. 1996) of polarized
optical emission, corresponding to the location of known radio knots.  The
first {\it HST} images of the 3C 66B jet were discussed by Macchetto et al.
(1991), and the jet has also been imaged with {\it ISO} (Tansley et al. 2000)
and {\it Chandra} (Hardcastle et al. 2001).   

A tentative detection of the southern counter-jet was claimed in $I$ band by
Fraix-Burnet (1997), based on ground-based imaging.  As shown in Figure 3,
while there are some objects in the region of the counter-jet corresponding to
the regions claimed by Fraix-Burnet (1997), they appear more likely to be point
sources (perhaps globular clusters) than extended counter-jet features. They do
not exhibit any significant polarization structure at the levels that we are
able to detect, and given the multitude of globular clusters in the field, this
seems a more likely explanation.  A more in depth analysis of the spectra of
these features will be presented in an individual paper on this object,
including several more bands from the {\it HST} archive.

The jet of 3C 66B rivals that of M87 (n.b., the jets of 3C 273 and 3C 403  have
roughly similar angular extents, but are only bright for $\lsim 50\%$ of their
extents) for the longest angular length optical jet feature known.   It also
rivals M87 for the sheer variety of structure:  while the majority of the
optical emission comes from a region that appears smooth at lower resolution
(often referred to as knots B and C), high-resolution {\it HST} images reveal a
number of rather sharp, abrupt structures on a variety of distance scales. 
While we see optical emission from the entire length of the jet, our  {\it HST}
polarimetry has high-enough signal to noise to image the polarization
morphology of 3C 66B's jet for just over $7''$.  

3C 66B shows a wide variation in fractional polarization, with a minimum at the
downstream end of knot B ($\sim 10\%$) and a maximum ($\sim 35\%$) at knot B's
upstream edge. Knot D displays a 90$^\circ$ rotation in MFPA across large
regions of its flux maximum, while a smaller rotation of the magnetic field is
seen in the broad, flux maximum region of knot B, with the field vectors
running parallel to the flux contours, at an angle of $\sim 45^\circ$ to the
jet propagation direction.  Knot E also shows a strong rotation in the 
MFPA, with magnetic field vectors approximately perpendicular to the jet and
parallel to the local flux contours.  These characteristics are classic markers
of a perpendicular bow shock.
Unlike several of the other jets in our sample (3C
78 and 3C 264, this paper; also M87's knot A-B region, as in Perlman et al.
1999), in 3C 66B's jet we do not see significant increases in fractional
polarization at the edges accompanying a near-constant MFPA in the
interior. Small regions of $90^\circ$ rotation in MFPA are seen near the
polarization minima, suggesting that superposition plays a part in what we see
from these regions. This combination of characteristics makes 3C 66B's
polarization morphology unique among our sample.

{\it 3C 78.} The radio source 3C 78 (Figure 4) is hosted by the E/S0 galaxy NGC
1218, at a redshift of $z=0.0289$.  The optical galaxy  has a face-on dust disk
(Sparks et al. 2000).  Its radio structure (Saikia et al. 1986, Unger et al.
1984, Jones, Sramek \& Terzian 1981) shows on the largest scales a two-sided
source with both jets surrounded by an edge-dimmed ``halo'' of emission.
The two jets lie at about 120$^\circ$ from one another on kpc scales,
suggesting a wide-angle tail morphology. Its 178 MHz radio power lies near the
FR I/FR II boundary, as commonly seen for wide-angle tail type sources. On
smaller scales, only the NE jet is seen, with bright emission extending about
$2''$ from the core. The optical emission from its jet, which extends for about
$1.5''$, was first discovered by Sparks et al. (1995). 

The radio and optical morphology of the 3C 78 jet is remarkably smooth, with
only a few bright knots. Our image shows polarized emissions averaging $\sim
12\%$ for the entire length of the optical jet. It also shows evidence of
oversubtraction in the inner regions, due to difficulty in fitting out the
surface brightness variations due to the dust lane. 3C 78 shows the least
variation in fractional polarization of all the jets in this sample.   We do,
however, observe lower polarizations near the jet center, and   maxima at the
jet edges (reaching as high as 50\%) and in interknot regions (15-25\%). Only
minor changes in  magnetic field position angle are seen, even in knot flux
maxima. The only significant changes that are seen occur in two regions: at
$0.1''$ from the nucleus (just outside the area where there is significant
residual from PSF  subtraction) the MFPA is perpendicular to the jet
direction, then  switches within $0.1''$ to a parallel configuration.  A more
subtle feature is seen at about $0.7''$ from the nucleus (near the end of the
brightest region of the jet), where the MFPA largely  follows flux
contours, around 30$^\circ$ away from the jet direction.

{\it 3C 264.} 3C 264 (Figure 5) is the second-nearest object discussed in this
paper, at a redshift $z=0.0217$. Its host galaxy is NGC 3862, an elliptical
galaxy in Abell 1367. The optical galaxy has a face-on dust ring at radii
between $0.75''$ and $1''$, as noted by Sparks et al. (2000) and also seen in
our data, and a relatively flat isophotal distribution inside these radii. The
galaxy's nuclear spectral energy distribution (SED) was found to be relatively
similar to those of BL Lac objects, suggesting that it might house a slightly
misaligned analog to those sources (Capetti et al. 2000, Rector et al. 2000).
3C 264 has a Fanaroff-Riley type I radio structure (e.g., Fanaroff \& Riley
1974) with a head-tail morphology. The radio jet extends $\sim 25''$ along a
direction that is initially towards the northeast, but bends towards the north
at $3-6''$ from the nucleus, and then back towards the northeast at greater
distances (Gavazzi, Perola \& Jaffe 1981; Bridle \& Vall\'ee 1981; Baum et al.
1988; Lara et al. 1999).

The optical jet of 3C 264 was first discovered by Crane et al. (1993) in
pre-COSTAR {\it HST/FOC} observations. The optically detected portion of the
jet in these data is considerably shorter than the radio jet: $\sim 2''.5$ long
versus $25''$, even though the Stokes I image we present here is deeper than
previous optical images. Lara et al. (1997) and Baum et al.  (1999) have
published comparisons of the jet's morphology in the optical and radio, as well
as the first information on its optical spectral index and optical emission
mechanism. The latter comparison was done on a jet-wide basis rather than for
individual jet components, but importantly, Baum et al. pinned down the
emission mechanism in the optical as synchrotron radiation, as seen in several
other optical jets. Radio polarimetry of the 3C 264 jet has recently been
discussed by Lara et al. (2004), using multi-frequency observations with the
{\it VLA} and {\it MERLIN} arrays in 1997 and 1999. The jet has also recently
been detected in the X-ray with {\it Chandra} (Padgett et al., in prep).

As mentioned, our Stokes I image is deeper than previous optical images of 3C
264, so we are able to describe the jet's morphology in greater detail.  The
morphology is quite smooth, and even though four knots (which we label A, B, C
and D in order of increasing distance from the core) are seen within the
innermost arcsecond, they all represent rather small enhancements in surface
brightness over the surrounding regions. Perhaps in line with its smooth
appearance, the polarization morphology of this jet is very homogenous,  with
few changes in MFPA, and with little apparent correlation between the
surface brightness and the few changes that we see. There is some trend to
see lower fractional optical polarization along the jet center line and
particularly at the positions of the knot maxima. By contrast, within the
innermost $1''$ (the optically brightest part) the ``edge'' is consistently a
polarization maximum. The jet undergoes a significant change in appearance and
morphology when it crosses the dust lane: as previously noted by Lara et al.
(1999), it fades precipitously, and then spreads out and appears to break up
beyond the lane, suggesting an interaction (even though the lane appears to be
seen essentially face-on, Sparks et al. 2000). Despite this, we see two
enhancements in optical surface brightness in or beyond the dust lane. Because
of the faintness of the jet in this region our signal to noise in $P$ is not
high.  However, we detect significant optical polarization beyond the dust lane
($\sim 45\%$) with MFPA aligned roughly parallel to the jet direction.

{\it 3C 346.} 3C 346 (Figure 6) is the most distant object discussed in this
paper, with a redshift $z=0.1610$ (e.g., Laing et al. 1983). It is a fairly
compact source, and was originally classified as a Compact Steep Spectrum (CSS)
object (Fanti et al. 1985). However, Spencer et al. (1991) argue that its
luminous core and small, one-sided distorted jet structure are best explained
by the foreshortening of a normal radio galaxy structure by a combination of a
relatively small angle to the line of sight, plus beaming. In that case, its
morphology and power would rank it as either a low-power FR II source or a
high-power FR I. Most subsequent authors have agreed with the foreshortened FR
II interpretation, and in particular Cotton et al. (1995) used the VLBI
radio-core dominance and jet to counterjet ratio to argue for a viewing angle
of $<32^\circ$ and a speed of $v=0.8c$. At low frequencies, this radio source
appears double-lobed, with a kinked jet that breaks up into a number of knots
at high ($>8$ GHz) frequencies (Cotton et al. 1995, van Breugel et al. 1992).

The optical host to this radio source is a 17th magnitude galaxy.  It has a
companion that is nearly as bright about $4''$ to the south. Optical emission
from its jet was first discovered by Dey \& van Breugel (1994), who found
excess emission in ground-based U and R images which they attributed to the
brightest two radio knots, called knots B and C by van Breugel et al. (1992).
{\it HST} snapshots (de Koff et al. 1996, de Vries et al. 1997) revealed an
excellent correspondence between the locations of optical and radio bright
knots, with optical emission seen from all regions of the radio jet. Neither of
the lobes is seen in the optical. X-ray emission from 3C 346 was first detected
with {\it Einstein} (Fabbiano et al. 1984).  ROSAT and ASCA observations
revealed that the source's X-ray emissions contained contributions both from a
power-law, presumably nuclear jet-dominated component, plus a thermal,
presumably cluster-related component. (Almudena Prieto 1996; Worrall \&
Hardcastle 1999, Worrall \& Birkinshaw 2001).  Worrall \& Birkinshaw (2001)
detected $\sim 3 \sigma$ evidence for variability of the unresolved, power-law
component.  More recently, {\it Chandra} observations (Worrall \& Birkinshaw
2005) show X-ray emission associated with (but offset somewhat from the flux
maximum of) the radio-optical knot C.

Our optical polarization map shows a wide variation in both degree and position
angle of polarization (Figure 6).  Knots B and D show somewhat reduced  optical
polarization percentages at their optical flux maxima  as compared to regions
just downstream.  However, knot C shows a more complex structure, with a
maximum in polarization at its southeast edge, near the location of the bend,
and a general gradient in polarization extending northwest to southeast.  We do
not have adequate signal to noise in most other knot regions. Published radio
polarization maps of 3C 346 (Akujor \& Garrington 1995, van Breugel et al.
1992) do not show either of these tendencies.  Also, the optical polarization
maps show a fairly complex morphology in MFPA.  In most parts of the
optical jet the magnetic field vectors track the flux contours and hence are
parallel to the jet direction.  However, a strong rotation of the MFPA is
observed in the knot C region, agreeing with the model postulated by Worrall \&
Birkinshaw (2005) that attributes the bend at this location to an oblique
shock, possibly associated with a wake caused by the passage of the companion
galaxy. We may also observe perpendicular optical MFPA both in knot D as
well as about 1'' downstream, near the terminus of the main optically-bright
region of the jet; however, the signal to noise is much
lower in that region.  .

{\it 3C 371.} 3C 371 (Figure 7) is a well-known object, first associated with the BL Lac
class by Miller (1975), and later included in the 1 Jy sample of BL Lacs by
Stickel et al. (1991).  It is among the nearest and brightest BL Lacs, with a
redshift of $z=0.0510$.  Its radio morphology consists of  two lobes and
a $25''$ long, one-sided radio jet (Wrobel \& Lind 1990, Akujor et al. 1994).
It has one of the highest jet to counterjet ratios known, 1700:1, implying a
viewing angle $\theta<18^\circ$ and a bulk Lorentz factor $\Gamma \geq 3.2$
(Gomez \& Marscher 2000).  However, despite frequent monitoring with the VLBA,
no superluminal motion has been detected, making it a bit of an oddity in the
BL Lac class.  Optical jet emission from 3C 371 was first detected in
ground-based images by Nilsson et al. (1997), and then confirmed in HST
observations by Scarpa et al. (1999).  Those authors noted optical emission
from three jet regions, specifically knots B ($1.7''$ from the nucleus), A
($3.1''$ from the nucleus) and D ($4.5''$ from the nucleus), and showed that in
the region of overlap, the optical and radio emission tracked one another
closely.  X-ray emission from the jet was first discovered by Pesce et al.
(2001), using {\it Chandra} observations, which detect knots A and B only.

Our Stokes I image is considerably deeper than the image published by Scarpa et
al., and clearly shows emission from the region between knots B and A. From
this we see that knot B appears to have a double structure, with a second
maximum about $0.6''$ downstream from the main one. Emission is also seen from
the inner $1.5''$ of the jet, although it is difficult to decipher its
morphology due to contamination from the diffraction spikes of the nuclear
point source (n.b., this issue was not discussed by Scarpa et al.,  although
the emission is clearly shown in their images - their Fig. 3). The additional
depth of our image allows us to rule out the possibility of a constant
radio-to-optical spectral index $\alpha_{ro}$ (Scarpa et al. 1999); we will
present more details about the spectral structure in an upcoming paper.

The polarization structure revealed by our data is complex. In the innermost
region of the jet (knot C), we detect polarized emission although the signal
there is somewhat affected by the diffraction spikes.  In this region the
polarization appears relatively constant in fraction as well as direction
(approximately perpendicular to the flux contours).  In knot B, we see a
polarization consistent with zero at the flux maximum, with only slight
rotations in the MFPA in the surrounding, moderately polarized  region. 
The most  complex structure, however, is seen in knot A and the interknot
region between knots A and B.  The A-B interknot region and the upstream end of
knot B display perpendicular magnetic fields and moderate polarizations. 
Immediately downstream of this, however, near the centerline of the jet  we see
a virtually unpolarized channel that is about $0.4''$ wide at the upstream end
of the knot and $0.2''$ wide at the downstream end.  Surrounding  this channel
on both sides are $\sim 0.5-1''$ wide regions where the polarization increases
smoothly and monotonically as one goes to the edge of the jet and the
MFPA generally follows the flux contours, with a  gradual rotation to a
perpendicular orientation at the downstream end.  Our signal to noise is not
high enough to detect polarized emissions from knot D.

\section{Polarization Properties of Jets}

With this paper, we have increased from 3 to 9 the number of jets for which
there exists optical/near-IR polarimetry with $\sim 0.1''$ resolution. Of the
other three jets that have been observed polarimetrically with HST, only one
(M87; see in particular Perlman et al. 1999, Capetti et al. 1997) has a signal
to noise that is comparable to most of the sources described here. This is
largely a byproduct of the surface brightness of the sources, as shown by Table
2, as well as the instrument configuration used.  The polarimetry observations
of the  remaining two sources, 3C 293 and 3C 273, have significantly lower S/N.
The S/N for the 3C 293 observations (Floyd et al. 2005) comes close to that
attained for the faintest jet regions discussed here, thanks to the fact that
NICMOS, rather than an optical instrument, was used for those observations (the
knots of the 3C 293 jet have SEDs that exhibit very sharp breaks at a few
microns [$R-K \approx 6$] both for intrinsic as well as extrinsic -- i.e.,
obscuration within the host galaxy -- reasons). However, in all but the
brightest regions, the S/N of the 3C 273 jet observations (Thomson et al. 1993)
is too low to be useful, and indeed its results are in serious conflict with
much higher S/N (albeit lower resolution) ground-based polarimetry (R\"oser et
al. 1996).

An examination of Table 2 shows that the sources with existing HST polarimetry
are strongly biased towards low powers.  Only three of the nine sources are
nominally of the FR II class, although a fourth has been called FR II by some
authors.  This observational bias is largely a byproduct of having observed the
optically brightest jets known in 1999, which were concentrated in relatively
low-z sources.  This  can be seen considerably better when one looks at the
total and jet radio powers at 5 GHz (data gathered from Liu \& Xie 2002).  We
show this in Figure 8, and plot in addition to these 9 sources the seven
highest optical surface brightness quasar and FR II jets (Sambruna et al. 2004,
Kraft et al. 2005). Figure 8 makes apparent the fact that, even with these
observations, there is still a large gap in our understanding of the magnetic
field structure of jets at optical frequencies, particularly at high
luminosities.  Further observations (to be requested in future cycles) are
needed to fill this gap.  

Even with the above proviso, however, it is still useful to take stock of what
we know about the polarization properties of jets in the optical.  What is
quite obvious from these data as well as previous observations is that  jets
display a great variety of optical polarization properties.  Most objects seem
to display at least some variation in either optical polarization  percentage
or magnetic field direction, although only small variations are seen in the
jets of 3C 66B (Figure 2), 3C 78 (Figure 4) and 3C 264 (Figure 5).  High
optical polarization seems to be quite typical, averaging $\sim 20\%$ in 
bright regions, and reaching values as high as  $50-70\%$ in  a few jet regions
(for example, M87, Perlman et al. 1999, 2003; some regions of the 3C 78 [Figure
4] and 3C 346 jets [Figure 6]).  All but one of these very high polarization
loci are well separated from their local flux maximum (the exception being knot
HST-1 of the M87 jet during its recent flare, Perlman et al. 2003).  However,
all of them seem to be in regions where there is some other indication of a
strong shock, as we discuss below for polarization maxima in general.  From
this, we can draw two fairly general inferences:  first, that the apparent
magnetic field orientation shown by these maps in optically emitting regions is
well ordered, and second, that the ordering of the magnetic field is
significantly affected by local, dynamical features within the jet flow.  The
first of these conclusions is similar to that based on radio polarization maps
of jets (e.g., Bridle \& Perley 1984).  It is important to emphasize that  what
these maps show us is the ordering and direction of the magnetic field in the
direction perpendicular to our line of sight.  This is indicative of  the
overall ordering and direction of the magnetic field, although components along
the line of sight are not traced by polarization data.  The effect of this last
point is dependent on the orientation of the jet and its various dynamical
features. (Laing 1980, 1981; Hughes, Aller \&  Aller 1985; Cawthorne \& Cobb
1990).  For example, in the cases of oblique shocks (Cawthorne 2006) or
anisotropic, disordered magnetic fields (Laing \& Bridle 2002; Canvin \& Laing
2004; Canvin et al. 2005), the combination of tangled magnetic fields with an
oblique line of sight can cause the polarization vector field to be an imprecise
indication of what the field directions are.

Regions of locally high polarization appear to be observed in two types of
regions within optical jets. When a region of high polarization is seen close
to a flux maximum and within the jet interior it is often, but not always,
associated with the region upstream of that flux maximum, with much lower
optical polarization being observed at the flux maximum. Several such regions
are seen in the M87 jet (Perlman et al. 1999), and it is also seen in some
other objects, for example knot A of the 3C 273 jet (Thomson et al. 1993), two
of the knots in the 3C 15 jet (Figure 1) and knot C of the 3C 371 jet (Figure
7).  As has been previously noted by Perlman et al. (1999), radio maps of most
jets do not show these trends as strongly. This is consistent with shocks at
these loci, where the local magnetic field is strongly compressed at the front.
The shock fronts are likely to be a site of particle acceleration, as in M87 
(Perlman \& Wilson 2005), where enhanced X-ray emission and harder spectra are
seen coincident or nearly coincident with polarization minima, suggesting that
at these loci the magnetic field might be highly disordered leaving more energy
available  for particle acceleration.  Because of the correlation with particle
acceleration, the optical polarization maps will show the  trend much more
clearly than the radio maps, which are far less sensitive to local jet
dynamics.   However, a full understanding of the interaction between the
magnetic fields in jets,  their dynamics and the resulting multiwaveband
emission will require an understanding of all scales.  We will return to this
subject in future papers, and we hope that these data will stimulate future
theoretical work.

 Locally high polarizations are also seen near the outer edges of the jet, the
vectors appear to follow the local flux contours, suggesting shearing along the
edge of the jet. Similar trends have been seen in radio polarization maps
(e.g., Bridle \& Perley 1984; Owen, Hardee \& Cornwell 1989).  This trend is 
particularly marked  in apparent ``working edges'' in regions where bends are
seen, as in, e.g., 3C 346, as well as 3C 15.  In these cases, the magnetic
field vectors appear to track the jet bend, indicating a strong link between
the local magnetic field and the dynamical factors that are bending the jet.
Regions in two other jets appear to have low polarization ``channels'' along
the centerline of the jet, either within one very large knot complex (as in the
case of the 3C 371 jet, see Figure 7) or following the ridgeline between
multiple knot maxima (as in the case of the 3C 78 jet, see Figure 4).  In these
regions, we could have interior regions with B perpendicular to the jet, thus
cancelling the polarization of the sheath (as originally suggested by Owen,
Hardee \& Cornwell 1989 to explain the radio polarization morphology in some
regions of the M87 jet),  or alternatively the magnetic field in these regions
is heavily tangled, a  model which makes rather less sense given the lack of
strong radio flux maxima.

\section{Summary}

We have presented {\it HST} optical imaging polarimetry for six optical jets,
thus increasing from 3 to 9 the list of objects for which data of this type
exist.  The objects seem to split into two classes, with three showing rather
simple magnetic field structures (3C 66B, 3C 78 and 3C 264), but more complex 
variations seen in the remainder. 

We have found a wide range of optical total flux and polarization morphologies,
which correlate with one another in a wide variety of ways.  We do, however,
see three trends in the data. As in the radio, we find that the fractional
polarization is often increased along the edges or most or all of these jets, 
with an MFPA that follows the jet direction.  However, somewhat differently 
from the radio, we find that optical flux maximum regions are often (but not
always) associated with local minima in fractional polarization, and that 
maxima in fractional polarization are usually well separated from the local
flux maxima.  We do not, however, find a universal or near-universal pattern. 
Variations in polarization are often, but not always, associated  with
rotations in MFPA.  We see changes in MFPA in many knot regions,
but not all.  Thus the patterns seen in Perlman et al. (1999) are not universal
to all jets, but we do see evidence that they are common to most
strong, shock-like features within jets.  

The picture that emerges for jets is therefore rather complex, and it is quite
likely that a great many factors could contribute to the general character of 
a given jet's magnetic field structure. There does appear to be a general
correlation between overall ``knottiness'' and the complexity of polarization
structure, and the most complex structures are  seen in regions that are either
shock-like or associated with apparent bends in the jet.  Thus we can say with
some certainty that the magnetic field in  jets is strongly affected by local
dynamics, and that this trend is seen much more prominently in optical maps. 
There is also a hint of a  correlation between the complexity of polarized
structure and jet power, suggesting some role for initial power as well as
perhaps jet speed in determining the overall nature of the magnetic field
structure in the jet as a whole.  However, it is important to stress that at
the present time this statement cannot be  extended to the higher powers
typical of FR II like objects due to the dearth of observations in
this power range.  Indeed, a full understanding of the interaction between
the magnetic fields in jets, and the relevant physical parameters over the full
range of jet properties, will require an understanding of the jet emission
on all scales, across the electromagnetic spectrum.  

\vfill\eject

\vfill\eject

\begin{figure}

\centerline{\includegraphics*[height=6in]{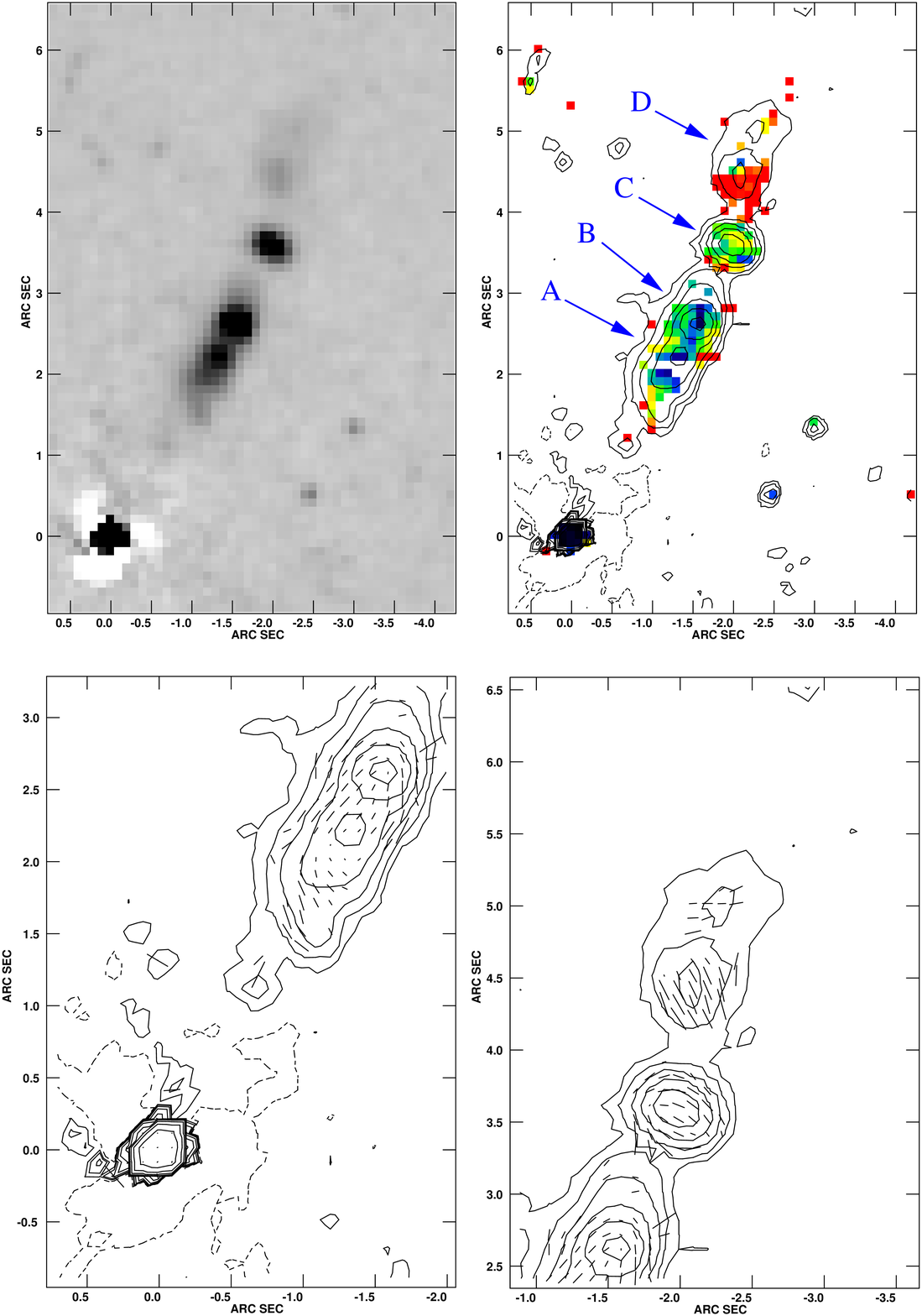}}

\caption{Optical polarimetry of the 3C 15 jet. The image scale is 1.39
kpc/arcsec.  At top left, Stokes I (total intensity), and at top right,
fractional polarization overlaid with contours of total intensity - red
indicates $\gsim 45\%$ polarization. At bottom left, contours of total
intensity superposed with polarization (apparent {\it B} field) vectors of the
inner jet region - knots A and B - , and at bottom right, the same for the
outer jet regions - knots C and D. A vector $1''$ long represents 500\%
polarization, and contours are spaced by powers of 2.  The image has been
galaxy subtracted (see \S 2 for details), and in all frames, north is up and
east is to the left.  The apparent negatives in flux surrounding the nucleus
are residuals from galaxy subtraction in the innermost regions; the likely
reason behind these features is nuclear dust. As can be seen, the 3C 15 jet
shows rich polarization structure, with field rotations in at least three
components and wide variations in the degree of polarization.  See \S 3 for
discussion.}

\end{figure}

\begin{figure}

\centerline{\includegraphics*[width=470pt]{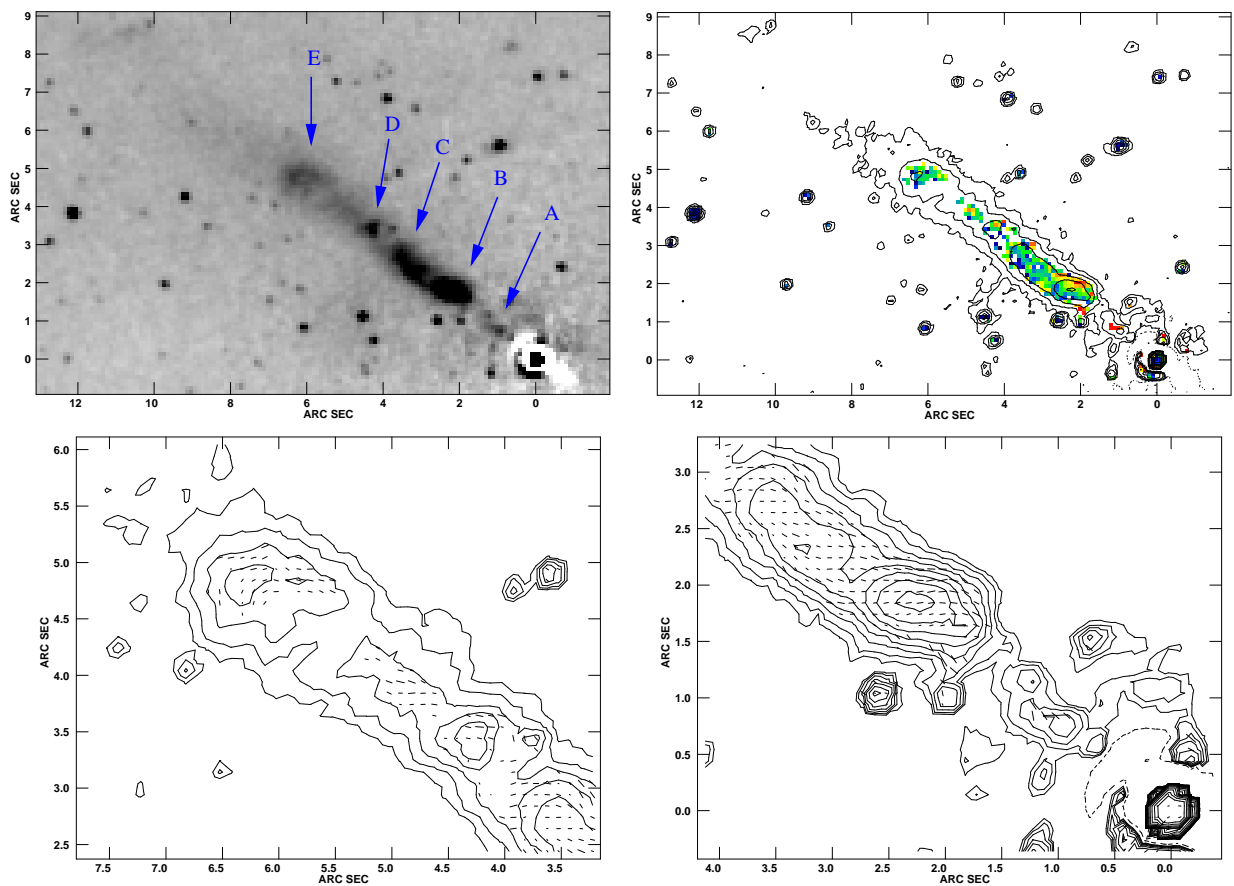}}

\caption{Optical polarimetry of the 3C 66B jet. The image scale is 440
pc/arcsec. The top left panel shows our Stokes I image with knots labelled
following Hardcastle et al. (2001).  At top right, we show fractional
polarization (colors) overplotted with total intensity contours - red indicates
$\gsim 30\%$ polarization. At bottom left, we show contours of total intensity
superposed with polarization (apparent {\it B} field) vectors for the outer
part of the jet - knots D and E -, and at bottom right we show the same for the
inner part of the jet - knots A, B, and C; a vector $1''$ long represents 400\%
polarization; and contours are spaced by powers of 2. The presence of a nuclear
dust ring (Sparks et al.  2000) can be seen in the residuals from galaxy
subtraction in the top left panel. Note the stability of the direction of the
magnetic field in the 3C 66B jet. See \S 3 for discussion.}

\end{figure}

\begin{figure}

\centerline{\includegraphics*[width=470pt]{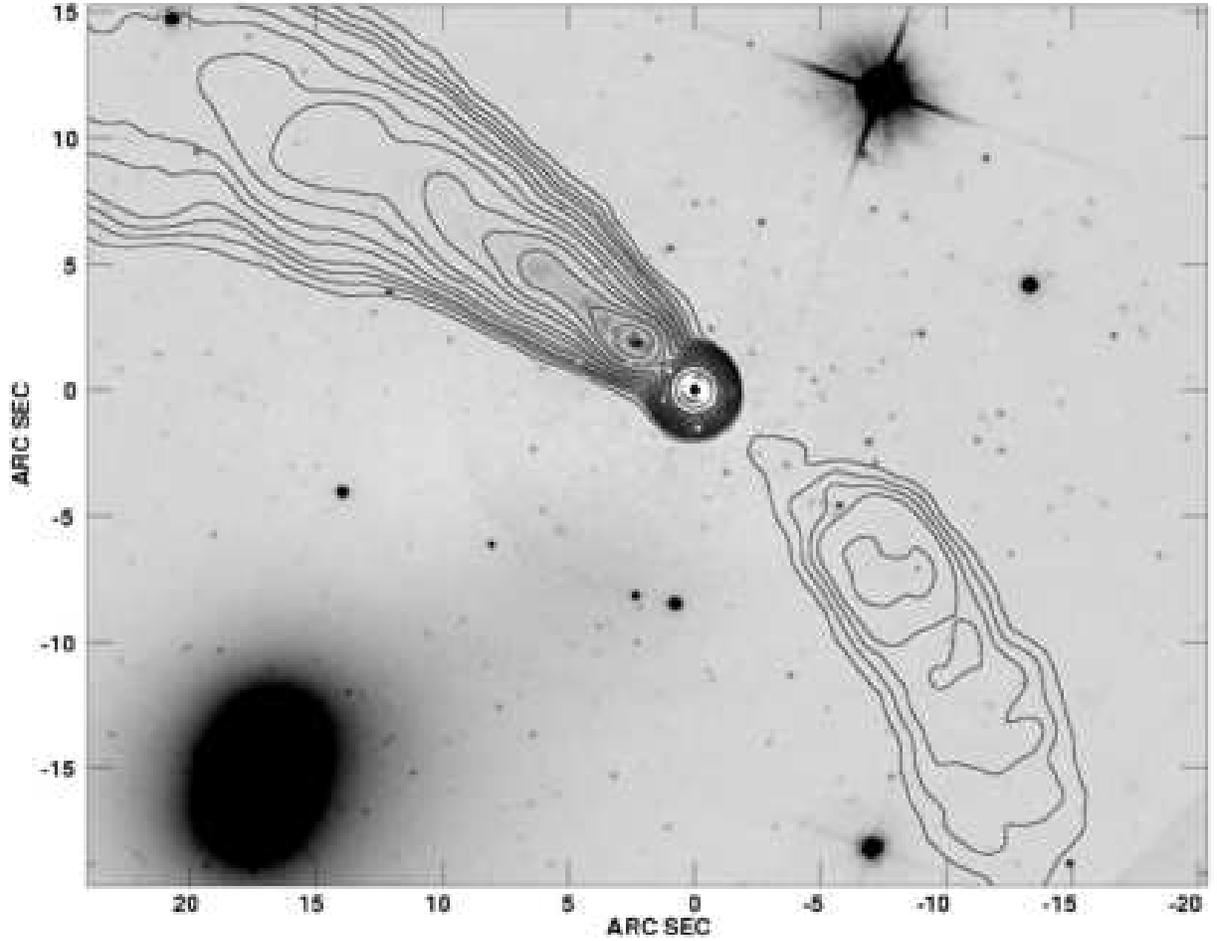}}

\caption{3C 66B Stokes I, galaxy subtracted image overplotted with radio
contours. This shows that we do not detect convincingly any counterjet
emission. In particular, note the large number of point sources in the region
of the counterjet, as well as over the entire field; many of these are likely
globular clusters associated with the host galaxy.}

\end{figure}

\begin{figure}

\centerline{\includegraphics*[width=470pt]{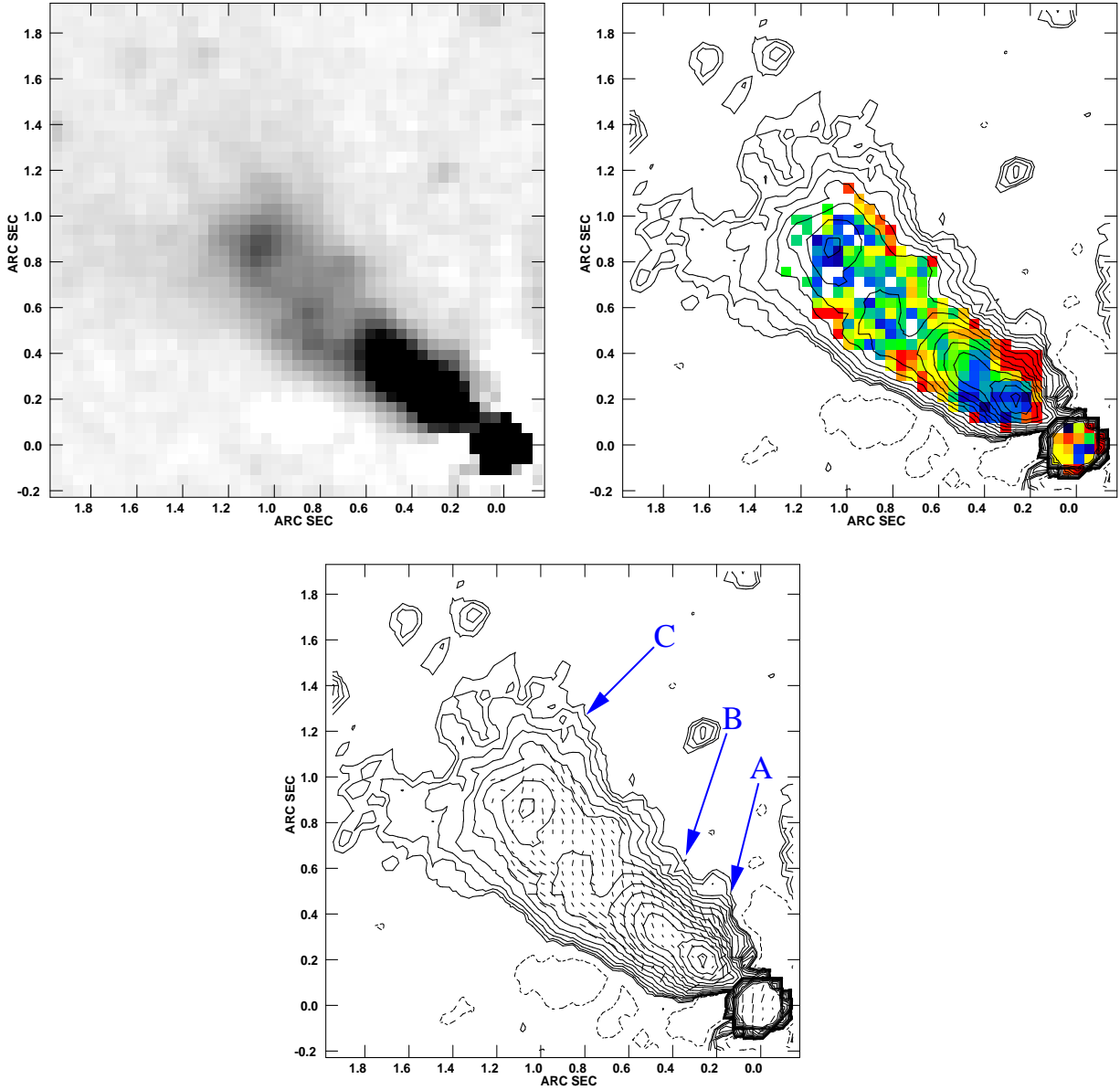}}

\caption{Optical polarimetry of the 3C 78 jet. The image scale is 600
pc/arcsec. The top left panel shows our Stokes I image. It has been scaled to
bring out the faint outer parts of the jet. At top right, we show fractional
polarization (colors) overlaid with contours of total flux - red indicates
$\gsim 40\%$ polarization - , while in the bottom panel, we show contours of
total intensity with polarization (apparent {\it B} field) vectors; a vector
$0.1''$ long represents 100\% polarization; and contours are spaced by
$\sqrt{2}$. The 3C 78 jet shows a wide variation in the degree of polarization,
but a relatively constant magnetic field direction, as can be seen.  All three
knots show reduced polarization at flux maximum. See \S 3 for discussion.}

\end{figure}

\begin{figure}

\centerline{\includegraphics*[width=440pt]{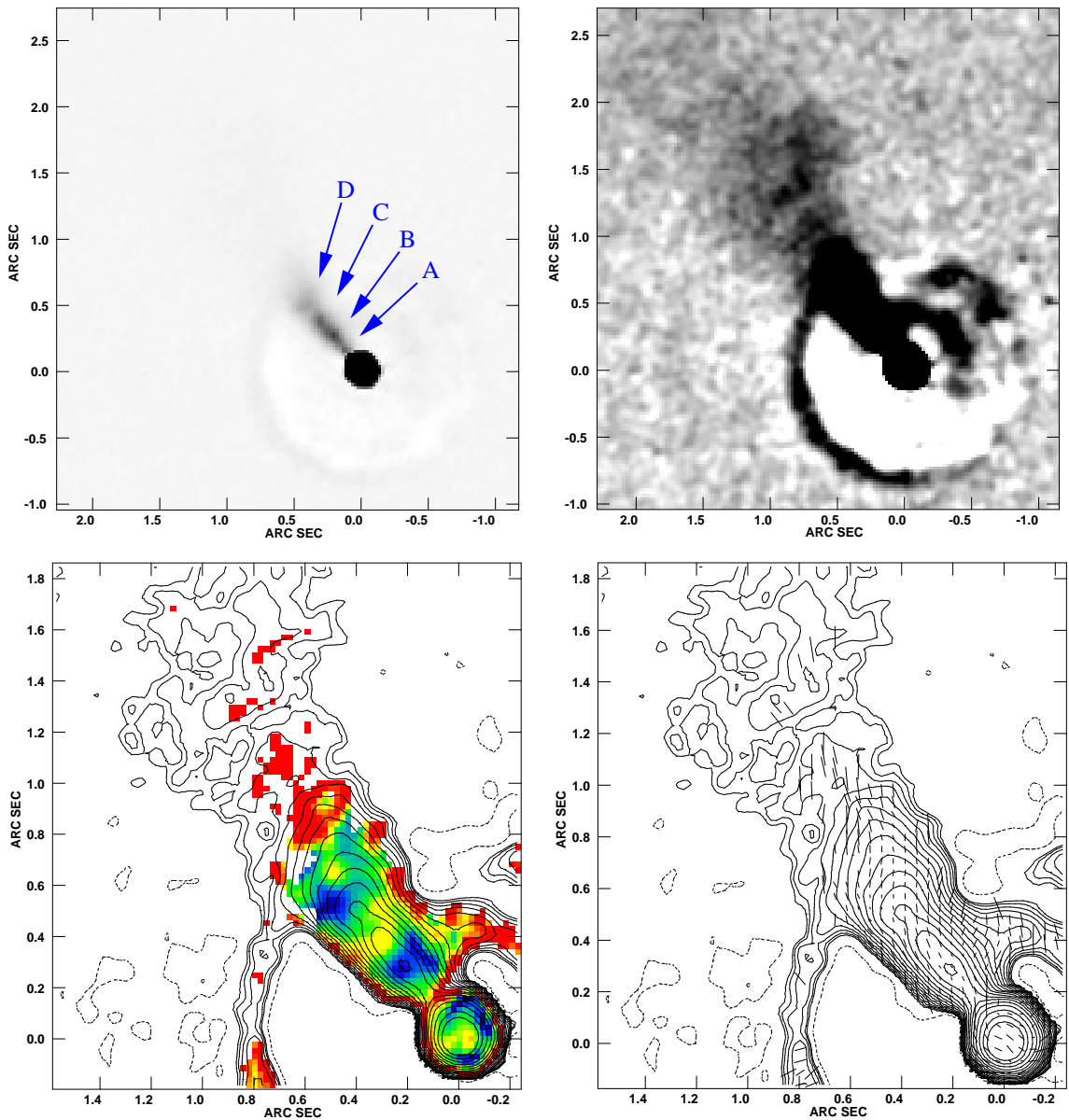}}

\caption{Optical polarimetry of the 3C 264 jet. The image scale is 440
pc/arcsec.  In the upper panels, we show two views of the Stokes I image.  The
top left panel shows a view designed to bring out fine-scale structure within
the knots, while the top right panel shows a stretch that saturates on bright
regions, but helps bring out faint structure in this jet.  Only in the view at
top right can the jet even be seen beyond the dust ring at $1''$ from the
nucleus (Sparks et al. 2000). At bottom left, we show fractional polarization
(colors) overlaid with contours of total flux - red indicates $\gsim 45\%$,
while at bottom right, we show contours of total intensity with polarization
(apparent {\it B} field) vectors; a vector $0.1''$ long represents 150\%
polarization; and contours are spaced by $\sqrt{2}$.  The 3C 264 jet shows
similar characteristics to the jet of 3C 78, with wide variation in the  degree
of polarization but a nearly constant magnetic field direction.  The
low-polarization regions appear to form a ``channel'' down the jet centerline.
See \S 3 for discussion.}

\end{figure}

\begin{figure}

\centerline{\includegraphics*[width=335pt]{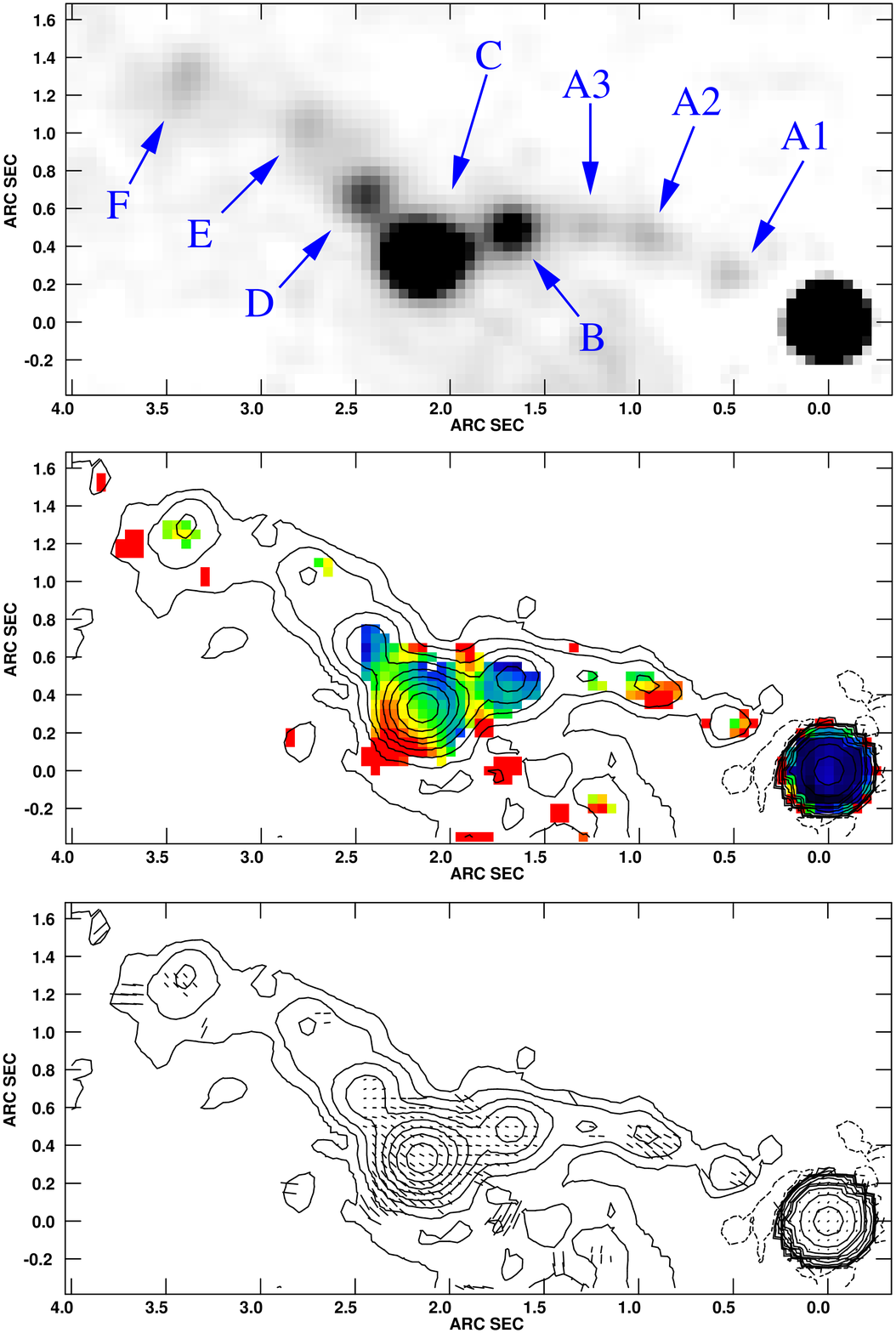}}

\caption{Optical polarimetry of the 3C 346 jet. At top, Stokes I (total
intensity), at middle, fractional polarization (colors) overlaid with contours
of total intensity - red indicates $\gsim 40\%$ polarization , and at bottom,
contours of total intensity superposed with polarization (apparent {\it B}
field) vectors; a vector $1''$ long represents 1000\% polarization; and
contours are spaced by powers of 2. The bright region to the south of the jet
is a residual from galaxy subtraction, due to the presence of a nearby galaxy.
In the 3C346 we do not see high polarizations in all jet components; some of
the faint knots are too faint to detect in polarized light. The bright knots
show significant variation in degree of polarization as well as magnetic field
direction, with a strong rotation, accompanied by increased polarization at the
location of the jet's prominent bend. See \S 3 for discussion.}

\end{figure}

\begin{figure}

\centerline{\includegraphics*[width=470pt]{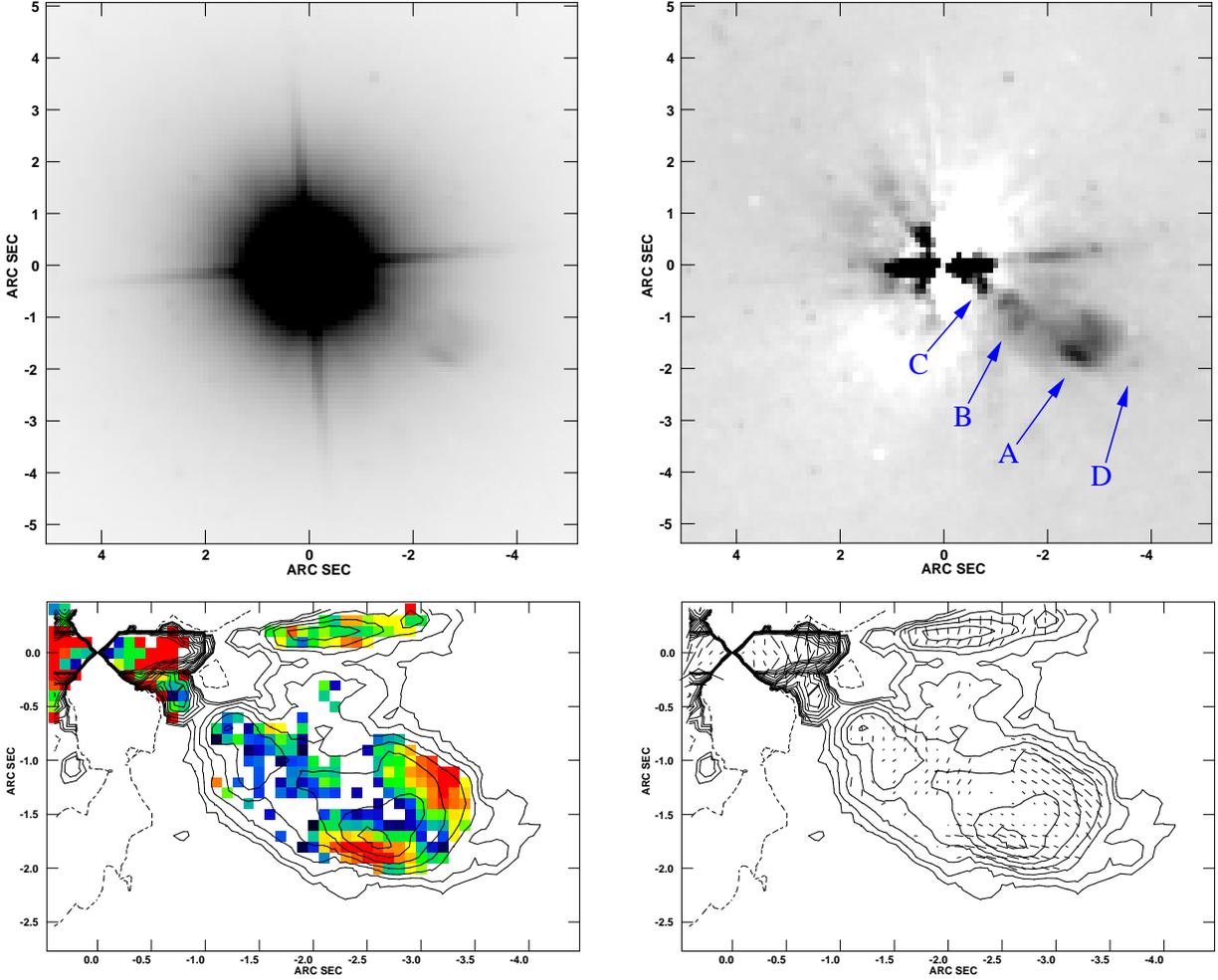}}

\caption{Optical polarimetry of the 3C 371 jet. At top left, Stokes I (total
intensity) without point source or galaxy subtraction, and at top right Stokes
I with our model subtracted. At bottom left, fractional polarization (colors)
overlaid with contours of total flux - red indicates $\gsim 30\%$ polarization.
At bottom right, contours of total intensity superposed with polarization 
(apparent {\it B} field) vectors; a vector $1''$ long represents 350\%
polarization; and contours are spaced by $\sqrt{2}$. The apparent negatives
surrounding the nucleus are residuals from galaxy subtraction, combined with
point-source mitigation techniques described in the text. Note the rich
polarization structure in this jet, with large variations in both degree and
direction of polarization. Perhaps most prominent  is the large,
low-polarization ``channel'' that appears to extend down the length of the
brightest jet component.  See \S 3 for discussion. }

\end{figure}

\begin{figure}

\centerline{\includegraphics*[scale=0.75]{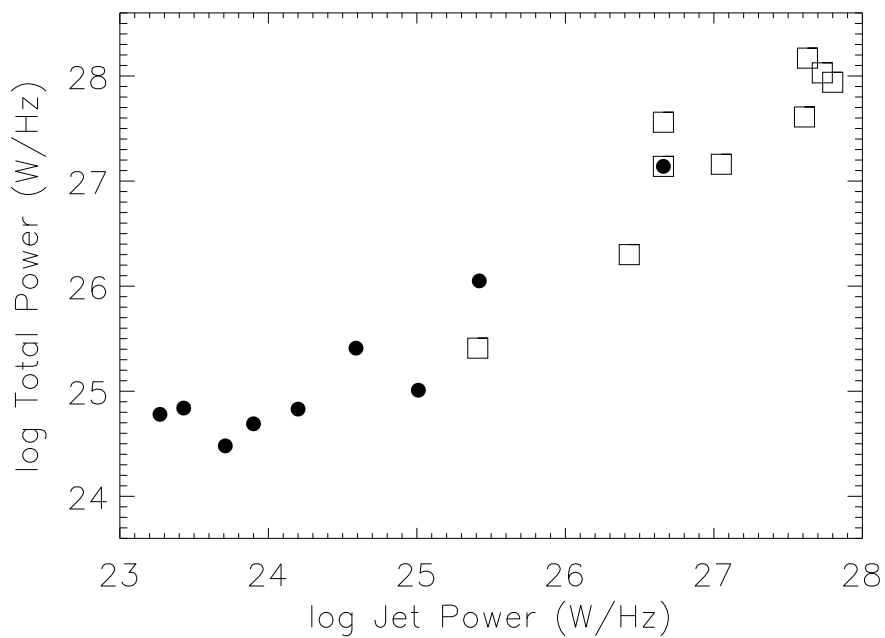}}

\caption{The radio jet power (at 5 GHz) and total power (at 1.4 GHz) of the
jets with {\it HST} polarimetry (data from Harris 2003, Jester 2003 and Liu \&
Zhang 2002), compared with the brightest Quasar and FR II jets (data from
Sambruna et al. 2004 and Kraft et al. 2005). Note that the jets discussed in
this paper are  clustered mainly at relatively low power, typical of FR I
sources, while in  only one high-power source, 3C 273, has the jet been
observed polarimetrically in  the optical, and its polarization data (Thomson
et al. 1993) are low S/N.}

\end{figure}

\end{document}